\documentclass[journal=jctcce,manuscript=article]{achemso}

\usepackage{amsmath}
\usepackage{amssymb}
\usepackage{amsfonts}
\usepackage{bm}
\usepackage{bbm}
\usepackage{braket}
\usepackage{mathtools}
\usepackage[Symbol]{upgreek}
\usepackage{hyperref}
\usepackage{microtype}
\usepackage{tikz}
\hypersetup{
    colorlinks=true,
    linkcolor=blue,
    citecolor=red,
    filecolor=magenta,      
    urlcolor=cyan
}

\DeclareMathOperator{\e}{\mathrm{e}}

\newcommand\nnfootnote[1]{%
  \begin{NoHyper}
  \renewcommand\thefootnote{}\footnote{#1}%
  \addtocounter{footnote}{-1}%
  \end{NoHyper}
}

\title{Complete Active Space Iterative Coupled Cluster Theory}

\setkeys{acs}{ maxauthors = 0 }

\author{Robin Feldmann}
\affiliation{ETH Z{\"u}rich, Department of Chemistry and Applied Biosciences, Vladimir-Prelog-Weg 2, 8093 Z{\"u}rich, Switzerland}
\author{Maximilian M\"orchen}
\affiliation{ETH Z{\"u}rich, Department of Chemistry and Applied Biosciences, Vladimir-Prelog-Weg 2, 8093 Z{\"u}rich, Switzerland}
\author{Jakub Lang}
\affiliation{Faculty of Chemistry, University of Warsaw\\
Pasteura 1, 02-093 Warsaw, Poland}
\author{Micha\l\ Lesiuk}
\affiliation{Faculty of Chemistry, University of Warsaw\\
Pasteura 1, 02-093 Warsaw, Poland}
\author{Markus Reiher}
\email{mreiher@ethz.ch}
\affiliation{ETH Z{\"u}rich, Department of Chemistry and Applied Biosciences, Vladimir-Prelog-Weg 2, 8093 Z{\"u}rich, Switzerland}

\date{September 4, 2024}

\begin{document}

\begin{abstract}

In this work, we investigate the possibility of improving multireference-driven coupled cluster (CC) approaches with an algorithm that iteratively combines complete active space (CAS) calculations with tailored CC and externally corrected CC. 
This is accomplished by establishing a feedback loop between the CC and CAS parts of a calculation through a similarity transformation of the Hamiltonian with those CC amplitudes that are not encompassed by the active space. We denote this approach as the complete active space iterative coupled cluster (CASiCC) ansatz. 
We investigate its efficiency and accuracy in the singles and doubles approximation by studying the prototypical molecules H$_4$, H$_8$, H$_2$O, and N$_2$.
Our results demonstrate that CASiCC systematically improves on the single-reference CCSD and the externally corrected CCSD methods across entire potential energy curves while retaining modest computational costs.
However, the tailored coupled cluster method shows superior performance in the strong correlation regime, suggesting that its accuracy is based on error compensation.
We find that the iterative version of externally corrected and tailored coupled cluster methods converge to the same results. 
\end{abstract}

\maketitle
\nnfootnote{
$^*$ Corresponding author.
}

\section{Introduction}
\label{sec:introduction}

A nagging issue in contemporary electronic structure theory is the notorious static-dynamic correlation problem. It will manifest itself if the Hartree--Fock (HF) determinant provides not just an inaccurate, but a qualitatively incorrect representation of the wave function. 
As a consequence, multiple determinants in a full configuration interaction (FCI) expansion of the electronic state possess similar non-negligible configuration interaction (CI) coefficients. In such cases, approximate methods that construct excitations systematically from a single reference determinant, such as coupled cluster with singles and doubles (CCSD), finite-order M{\o}ller--Plesset perturbation theory, CI with singles and doubles, yield either inaccurate or even completely unreliable results \cite{Helgaker2014_Molecular,Bartlett2009_Book}. 

Then, multireference methods are required that rely on selecting a model space built from a subset of determinants that have significant weights in the FCI expansion. 
If the model space is constructed from all possible excitations from a set of occupied orbitals into a set of unoccupied (or virtual) orbitals, the approach will be referred to as a complete active space (CAS) method\cite{Roos1987_CompleteActive} 
(cf. also the first-order reaction space approach\cite{Ruedenberg1982_FORS}).
The prevailing strategy for tackling multireference problems currently hinges on a self-consistent field (SCF) optimization of the orbitals and CI-coefficients of the CAS configuration interaction (CASCI) wave function, called the CASSCF method \cite{Roos1980_CASSCF}. 
Since exactly diagonalizing the Hamiltonian in the CAS approach scales factorially with the number of electrons and orbitals, approximate schemes were developed. Examples are perturbatively selected CI approaches\cite{Malrieu1973_SelectedCi1,Malrieu1983_SelectedCi2,Giner2013_SCI,Evangelista2014_AdaptiveCI,Holmes2016_HBCI1,Liu2016_ICI,Tubman2016_DeterministicSCI,Smith2017_HBCI2,Garniron2017_HybridSCI},
the density matrix renormalization group (DMRG) \cite{White1992_DMRG1,white1993_DMRG2,Chan2008_Review1,Chan2009_Review2,Marti2010_Review-DMRG,Schollwock2011_MPSReview,Chan2011d_DMRG3,Wouters2014_ReviewDMRG,Kurashige2014_ReviewDMRG,Olivares2015_DMRGInPractice,Szalay2015_TensorReview,Yanai2015_DMRGReview,Baiardi2020_Perspective}, FCI quantum Monte Carlo\cite{Booth2009_FCIQMC1,Cleland2011_FCIQMC2}, and many-body expanded FCI\cite{Eriksen2017_MbeFci}. For an overview 
of modern approximate FCI methods and their comparison see Ref.~\citenum{Eriksen2020_BenzeneGroundState}.

Following the diagonalization of the Hamiltonian in the CAS approach, one then has to consider those orbitals neglected by the CAS in the first place.
Otherwise, dynamic correlation would be lacking, which would lead to inaccurate energies. Dynamic correlation is often accounted for using multireference perturbation theory (MRPT). There is some freedom in choosing the zeroth-order Hamiltonian\cite{Lindh2020_DynamicsBook}
resulting in different flavors of MRPT as in CAS second-order\cite{Andersson1992_CASPT2} perturbation theory (CASPT2) or \textit{N}-electron valence state second-order perturbation theory (NEVPT2)\cite{Angeli2001_NEVPT2}. However, MRPT comes with significant drawbacks: (i) perturbation theory cannot deliver high accuracy\cite{Loos2022_Caspt2Nevpt2,Loos2022_Caspt3}, (ii) the size of the CAS is constrained by the substantial computational demands associated with the evaluation of many-body reduced density matrices (RDMs), and (iii) because the size of the CAS is limited, it may not always possible to include all important determinants in the CAS, which gives rise to the intruder state problem that results in considerable numerical instabilities.

One possibility to improve upon the accuracy of MRPT is the multireference CI (MRCI)\cite{Buenker1974_SelectedMRCI1,Meyer1977_icMRCI1,Shavitt1977_CI,Siegbahn1980_icMRCI2,Werner1982_icMRCI3,Buenker1983_SelectedMRCI2,Werner1988_icMRCI,Shavitt1998_ucMRCI,Shamasundar2011_NewIcMRCI,Szalay2012_MrciReview,Lischka2018_MrCiAndMrCcReview} 
approach, where the dynamic correlation is accounted for by diagonalizing the Hamiltonian in the basis of determinants generated from excitations from a multideterminantal reference. 
However, its considerable computational demand (a consequence of the high-order RDMs required) restricts its applicability to small systems.

For single-reference systems, the coupled cluster (CC) method is the state of the art, especially in the form of the CCSD model with an additional noniterative, perturbative triple component, known as CCSD(T)\cite{Raghavachari1989_CCSD-t-1,Bartlett1990_CCSD-t-2}. In parallel to the development of single-reference CC methods, there has also been an active development of multireference coupled cluster (MRCC) approaches\cite{
Jeziorski1981_JMMMRCC,Banerjee1981_icMRCC,Mukherjee1975_icMRCC,Mukherjee1997_NormalOrdering,Mavsik1998_BWMRCC,Piecuch1992_HilberSpaceCC,Meller1996_StateSpecificMrCC1,Mahapatra1998_StateSpecificMrCC2,Mahapatra1998_MkMRCC1,Mahapatra1999_MkMRCC2,Hanrath2005_MRexpT,Hanrath2006_MRexpT,Mahapatra2010_SingleRootMrCC,Maitra2012_MkMRCC3,Lyakh2012_MrccReview,Evangelista2018_MrccPerspective,Lischka2018_MrCiAndMrCcReview}. However, the transition from single reference CC (SRCC) to MRCC proved to be far from straightforward and a multitude of issues surfaced. Some examples are the multiple-parentage problem that creates
redundancies, intruder states, and nonterminating commutator expansions\cite{Lyakh2012_MrccReview,Evangelista2018_MrccPerspective,Lischka2018_MrCiAndMrCcReview}. These obstacles have not been overcome entirely, and consequently, no generally applicable MRCC method has emerged yet\cite{Lyakh2012_MrccReview}. 

Multireference-driven SRCC methods offer a practical alternative to the genuine MRCC methods.
These methods were first introduced by Oliphant and Adamowicz\cite{Oliphant1991_ExternallyCorrected0,Oliphant1992_ExternallyCorrected1}.
Subsequently Piecuch and the former authors developed the CCSDt and CCSDtq methods\cite{Piecuch1993_SplitAmplitude1,Piecuch1999_CCSDt}(formerly known as state-selective (SS) MRCCSD(T) and SSMRCCSD(TQ)), which optimize the single and double amplitudes with a subset of the triple amplitudes and a subset of triple and quadruple amplitudes, respectively. 
In order to formalize these approaches, the authors developed the split-amplitude Ansatz \cite{Piecuch1993_SplitAmplitude1,Piecuch1994_SplitAmplitude2_H8,Piecuch1994_SplitAmplitude3_H8,Piecuch1995_SplitAmplitude4_H2O}, 
which builds the foundation of many modern MR-driven SRCC methods.
Following this work, Paldus and colleagues\cite{Paldus1997_ExternallyCorrected1,Paldus1998_ExternallyCorrected2,Paldus1994_ExternallyCorrected3,Paldus1994_ExternallyCorrected4,Paluds1997_ecCCSD-2}, and Stolarczyk\cite{Stolarczyk1994_ExternallyCorrected}, account for multireference effects by keeping MRCI-derived cluster amplitudes fixed while optimizing the remaining amplitudes within the SRCC framework.
The fixed amplitudes, known as internal amplitudes, are associated with excitations within the model (or internal) space, typically the CAS. The amplitudes related to excitations within the external space---composed of all other orbitals---are denoted external amplitudes. This ansatz was termed the spit-amplitude ansatz\cite{Piecuch1993_SplitAmplitude1}.
A particularly practical variant of the multireference-driven methods is the tailored CC (TCC) method, introduced by Kinoshita, Hino, and Bartlett\cite{Kinoshita2005_TailoredCC1} which has been extensively investigated in the literature\cite{Hino2006_TailoredCC2,Veis2016_TCC-DMRG,Faulstich2019_TCCAnalysis,Pittner2019_LocalTCC,Pittner2020_LocalTCC,Faulstich2019_TCCNumerical,Kats2020_FCIQMC_TDCC1,Brandejs2020_RelativisticTCC,Boguslawski2021_TCC,Kats2022_FCIQMC_TDCC2,Morchen2022_TCC}. 
Computationally less appealing, but well explored is the externally corrected CC (ecCC) with singles and doubles (ecCCSD) method \cite{Paldus1982_ecCCSD-1,Paluds1997_ecCCSD-2,Piecuch2018_ecCCSD-1,Piecuch2021-ecCCSD-2,Chan2021_ecCCSD-DMRG,Pittner2023_HilbertSpaceTCC,Bartlett2023_pCCD-TCC}.
Additionally, we mention the complete active space coupled cluster method\cite{Adamowicz2000_cascc1,Ivanov2000_cascc2,Lyakh2005_cascc3} which factorizes the wave function as a product of an exponential ansatz for the  external amplitudes and a linear CI ansatz for the internal CI coefficients, and the amplitudes and coefficients are optimized simultaneously via the projected residual equations.
We would also like to point out the method-of-moments CC (MM-CC) -- and a special case of this ansatz, the completely renormalized CC (CR-CC) -- which were developed in order to accurately describe bond breaking situations within the CC framework.\cite{piecuch2004,kowalski2000,kowalski2000a}
Later, the CC(P;Q)\cite{shen2012,shen2012a,shen2012b,bauman2017,deustua2017} ansatz emerged as a biorthogonal generalization of the MM-CC method.
However, a severe limitation of multireference-driven methods is the choice of the reference determinant. This limitation makes, e.g., CCSD more accurate than TCCSD within the single-reference regime due to the absence of back-coupling between the external and internal amplitudes and reduces the accuracy in the multireference regime due to the bias of the reference determinant.\cite{Faulstich2019_TCCNumerical,Faulstich2019_TCCAnalysis,Morchen2022_TCC}

While CC theory is often seen as a nonlinear wave function formulation, it can also be viewed as an effective Hamiltonian theory.\cite{Primas1963_EffectiveHamil}
This interpretation establishes a connection between CC and the similarity renormalization group (SRG) approach, which was developed by G{\l}azek and Wilson\cite{Wilson1993_SRG1} and by Wegner\cite{Wegner1994_SRG2}. 
The SRG method relies on a flow equation for unitary transformation of the Hamiltonian. This equation is integrated with a suitably chosen generator for the transformation, such that the off-diagonal elements of the Hamiltonian are vanishing.
This idea was later applied in chemistry by White \cite{White2002_SRG-CanonicalTransformation} who extended the SRG method to transformations in the many-body space.
This flavor of the SRG has also found numerous applications in nuclear structure theory where it is usually denoted as the in-medium SRG (IMSRG). \cite{Tsukiyama2011_IMSRG1,Tsukiyama2012_IMSRG2,Hergert2016_IMSRG-Rev} The IMSRG is a relatively new method that has seen rapid development in recent years.\cite{Hergert2016_IMSRG-MR,Gebrerufael2017_MRIMSRG-NCSM,Tichai2022_CombiningDmrgImsrg} 

Based on the idea of the IMSRG and White's canonical transformation, Evangelista developed the driven SRG (DSRG) method and applied it to chemical problems.\cite{Evangelista2014_DSRG1,Evangelista2019_DSRG-Rev} The advantage of the DSRG is that it does not require the integration of the flow equation, instead it can be solved in one step. Later, the DSRG was also generalized to a multireference method \cite{Evangelista2015_MRDSRG} based on the Mukherjee--Kutzelnigg normal ordering\cite{Mukherjee1997_NormalOrdering,Kutzelnigg1997_NormalOrdering}. 
Moreover, based on the DSRG, Li and Evangelista developed a new MRPT which alleviates the intruder state problem.\cite{Evangelista2015_MRDSRG,Evangelista2017_DSRG_PT3} 
However, a notable drawback of the DSRG method is its dependency on the flow parameter. The choice of this parameter significantly affects the energy and, if applied within the MRPT formulation, it also affects the intruder state problem \cite{Evangelista2015_MRDSRG}.

This work was inspired by the IMSRG approach, and we build upon the TCCSD and the ecCCSD (with three- and four-body amplitudes from the CAS) methods, and also on recent advances in SRCC theory, namely the subalgebra formulation of CC \cite{Kowalski2018_Subalgebra1,Bauman2019_Subalgebra2,Bauman2022_Subalgebra3,Kowalski2023_Subalgebra4} which, in turn, is based on the split-amplitude ansatz\cite{Piecuch1993_SplitAmplitude1,Piecuch1994_SplitAmplitude2_H8}.
Our approach starts from a standard CASCI plus TCCSD (or ecCCSD) calculation. 
We investigated the possibility of iteratively improving the TCCSD or ecCCSD wave function by applying a similarity transformation of the Hamiltonian with the CC amplitudes from the external space.
The many-body expansion of the similarity-transformed Hamiltonian is then truncated after a given order, as in the IMSRG method. In contrast to the IMSRG, the CC transformation results in a nonhermitian Hamiltonian that is subsequently projected into the CAS many-body space and then diagonalized with a nonhermitian FCI solver. 
Building on the work of Kowalski, who demonstrated that an iterative solution of the CC equations in two separate subalgebras is equivalent to solving the CC equations in the entire space \cite{Kowalski2018_Subalgebra1}, we reapply the TCCSD (or ecCCSD) approach.
The Hamiltonian is then subjected to another transformation and is once again diagonalized. 
This process is iteratively repeated until convergence is attained.
Our reasoning is the following: with this strategy we aim to address the challenge of the multireference-driven CC methods, where issues arise due to a lack of feedback between the amplitudes associated with the CAS and the external space. 
We note that Li and Evangelista have developed an iterative method based on the DSRG\cite{Evangelista2016_RelaxedDSRG,Evangelista2020_ldsrg3}, Liao, Ding, and Schilling have devised an iterative quantum-information theory informed TCC algorithm\cite{Liao2024_IterativeTCC}, and further ideas for the CC method have been proposed in the literature before\cite{Piecuch1998_Review,Chan2005_NO-DMRG,Lyakh2012_MrccReview,Kvaal2022_ThreeLagrangians}, first mentioned by Adamowicz, Piecuch, and Ghose, to the best of our knowledge. However, we are not aware of any practical realization of these ideas.

This work is organized as follows: in Sec.\ \ref{sec:theory} we provide the derivation of the working equations of the complete active space with iterative coupled cluster and we present a detailed explanation of the algorithm and its implementation. In Sec.\ \ref{sec:comp_details} we describe our computational methodology and we continue in Sec.\ \ref{sec:application} by presenting the computational results for several small molecules, namely H$_4$, H$_8$, H$_2$O, and N$_2$.

\section{Theory}
\label{sec:theory}

\subsection{The coupled cluster ansatz and its excitation subalgebras}
\label{sec:srcc}

In this section, we briefly review the single reference coupled cluster ansatz and introduce the excitation subalgebras by Kowalski\cite{Kowalski2018_Subalgebra1} to define the CC equations in the internal space, which will correspond to the CAS, and the external space, which consists of all other excitations that are not contained in the CAS. 
We highlight here that the subalgebra approach with the split-amplitude ansatz is based on the original work by Piecuch, Oliphant, and Adamowicz\cite{Piecuch1993_SplitAmplitude1,Piecuch1994_SplitAmplitude2_H8}. Especially, Ref.~\citenum{Piecuch1998_Review} reviews this formalism with a linear parametrization for the CASCI wave function and an exponential ansatz for the external CC wave function.
This framework enables us to derive our iterative optimization algorithm, where the internal equations can be solved with any nonhermitian CASCI method, and the amplitudes extracted from the CASCI wave function are either inserted in the TCC or in the ecCC ansatz.

We introduce the nonrelativistic electronic Born--Oppenheimer Hamiltonian in second quantization
\begin{equation}
    H = E_0 + \sum_{pq} F_{pq}  \{p^\dagger q\} + \frac{1}{4} \sum_{pqrs} \langle pq||rs\rangle \{p^\dagger q^\dagger s r\},
\end{equation}
where $F_{pq}$ is the Fock matrix and $\langle pq||rs\rangle$ 
are the antisymmetrized two-electron integrals in the spin-orbital basis. The latter are given in the physics notation.
In this work, we follow the standard convention where $p,q,r,s$ correspond to arbitrary spin orbitals, $i,j,k,l,\dots$ to occupied spin orbitals, and $a,b,c,d,\dots$ to unoccupied ones. Curly brackets refer to a normal-ordered string of standard Fermionic second quantization operators, where the normal-ordering is carried out with respect to the Fermi vacuum.
The full coupled cluster (FCC) wave function ansatz is an exact ansatz to solve the electronic Schr\"odinger equation and can be written as
\begin{equation}
    | \Psi_\mathrm{FCC} \rangle = \e^{T} |0\rangle,
\end{equation}
where $|0\rangle$ is the reference determinant and $T$ is the cluster operator defined as
\begin{equation}
    T = \sum_{n=1}^N T_n.
\end{equation}
Here, $N$ is the number of electrons in the system and $T_n$ contains $n$-body excitations and reads
\begin{equation}
    T_n = \frac{1}{(n!)^2} \sum_{\substack{i_1\dots i_n\\ a_1\dots a_n}} t_{i_1\dots i_n}^{a_1\dots a_n} \{ a^\dagger_1 \dots a^\dagger_n i_n \dots i_1 \},
\end{equation}
where, $ t_{i_1\dots i_n}^{a_1\dots a_n}$, are the CC amplitudes which we aim to optimize. Equivalently, the FCI wave function, 
\begin{equation}
    | \Psi_\mathrm{FCI} \rangle = (1+C)|0\rangle,
\end{equation}
also solves the Schr\"odinger equation exactly
with the CI operator
\begin{equation}
    C = \sum_{n=1}^N C_n,
\end{equation}
and the $n$-body CI excitation operators
\begin{equation}
    C_n = \frac{1}{(n!)^2} \sum_{\substack{i_1\dots i_n\\ a_1 \dots a_n}} c_{i_1\dots i_n}^{a_1\dots a_n} \{ a^\dagger_1 \dots a^\dagger_n i_n \dots i_1 \}.
\end{equation}
By cluster analysis\cite{Monkhorst1977_ClusterAnalysis}, a mapping between the CI coefficients, $c_{i_1\dots i_n}^{a_1\dots a_n}$, and the CC amplitudes, $ t_{i_1\dots i_n}^{a_1\dots a_n}$, can be found.
The associated determinant for a given string of second quantization operators acting on the reference is abbreviated as
\begin{equation}
   \ket{_{i_1\dots i_n}^{a_1\dots a_n}} = \{a^\dagger_1 \dots a^\dagger_n i_n \dots i_1\} | 0 \rangle.
\end{equation}
To formulate the equation for the CC amplitude optimization in a convenient way, we introduce projection operators. The projector onto the reference determinant is defined as
\begin{equation}
    P = \ket{0}\bra{0},
\end{equation}
and the projector onto all excited determinants is given by
\begin{equation}
    Q = \sum_{n=1}^N \sum_{\substack{i_1<\dots<i_n\\ a_1<\dots<a_n}} \ket{_{i_1\dots i_n}^{a_1\dots a_n}}\bra{_{i_1\dots i_n}^{a_1\dots a_n}}.
    \label{eq:virtual_projector}
\end{equation}
The energy-independent CC equations for the amplitudes are written as
\begin{equation}
    Q \e^{-T}H \e^{T} \ket{0} = Q (H \e^{T})_C \ket{0} = 0,
    \label{eq:cc_amplitude_eqns}
\end{equation}
where the subscript $C$ means that only the connected diagrams are to be considered. By solving Eq.~(\ref{eq:cc_amplitude_eqns}), the CC amplitudes, $t_{i_1\dots i_n}^{a_1\dots a_n}$, can be obtained, and the energy can be evaluated according to
\begin{equation}
    E = \bra{0}(H \e^{T})_C \ket{0}.
\end{equation}
Next, we divide the space of excitations into the internal excitations which denote the excitations associated with distributing $K$ electrons among $L$ orbitals often denoted as CAS($K$,$L$), and into the external space, which contains all other determinants.
We introduce the internal cluster operator $T^\mathrm{int}$, which generates the CASCI wave function in this space as
\begin{equation}
    |\Psi_{\mathrm{int}}\rangle = \e^{T^\mathrm{int}} \ket{0}.
\end{equation}
Additionally, we introduce the external cluster operator, $T^\mathrm{ext}$, which generates all other excitations. The FCC wave function can then be expressed as\cite{Piecuch1993_SplitAmplitude1}
\begin{equation}
    \ket{\Psi_\mathrm{FCC}} =  \e^{T^\mathrm{ext} + T^\mathrm{int}} \ket{0} = \e^{T^\mathrm{ext}} |\Psi_{\mathrm{int}}\rangle.
\end{equation}
The external and internal spaces can be viewed as the subalgebras introduced by Kowalski\cite{Kowalski2018_Subalgebra1} which enables us to rewrite the CC equations in a form that is suitable for our purpose. We rewrite the projection operator onto the excited-determinant manifold of Eq.~(\ref{eq:virtual_projector}) as
\begin{equation}
    Q = Q^\mathrm{int} + Q^\mathrm{ext},
\end{equation}
where $Q^\mathrm{int}$ projects onto all excited determinants from the CAS($K$, $L$), and $Q^\mathrm{ext}$ projects onto the external determinants.
Consequently, we can divide the Schr\"odinger equation into an internal part
\begin{equation}
    (P+Q^\mathrm{int}) H \e^{T^\mathrm{ext}} |\Psi_{\mathrm{int}}\rangle = E (P+Q^\mathrm{int})  \e^{T^\mathrm{ext}} |\Psi_{\mathrm{int}}\rangle,
    \label{eq:int_SE_1}
\end{equation}
and an external one
\begin{equation}
    Q^\mathrm{ext} H  \e^{T^\mathrm{ext} + T^\mathrm{int}} \ket{0} = E Q^\mathrm{ext}   \e^{T^\mathrm{ext} + T^\mathrm{int}} \ket{0}.
    \label{eq:ext_SE}
\end{equation}
We introduce the similarity-transformed Hamiltonian
\begin{equation}
    \Bar{H}^\mathrm{ext} = \e^{-T^\mathrm{ext}} H \e^{T^\mathrm{ext}},
\end{equation}
and insert the identity $\e^{T^\mathrm{ext}}\e^{-T^\mathrm{ext}}$ in front of the Hamiltonian in Eq,~(\ref{eq:int_SE_1}), such that we obtain
\begin{equation}
    (P+Q^\mathrm{int}) \e^{T^\mathrm{ext}} \Bar{H}^\mathrm{ext} |\Psi_{\mathrm{int}}\rangle =  E (P+Q^\mathrm{int}) \e^{T^\mathrm{ext}} |\Psi_{\mathrm{int}}\rangle.
        \label{eq:int_SE_2}
\end{equation}
Kowalski has demonstrated in Ref.\citenum{Kowalski2018_Subalgebra1} that $\e^{T^\mathrm{ext}}$ contributes neither to the left nor to the right-hand side of Eq.(\ref{eq:int_SE_2}), and hence, the equation can be simplified as
\begin{equation}
    (P+Q^\mathrm{int}) \Bar{H}^\mathrm{ext} |\Psi_{\mathrm{int}}\rangle =  E (P+Q^\mathrm{int})  |\Psi_{\mathrm{int}}\rangle.
        \label{eq:int_SE_fin}
\end{equation}
In this form, it becomes apparent that the amplitudes $T^\mathrm{int}$ can be obtained by solving the nonhermitian eigenvalue problem of Eq.~(\ref{eq:int_SE_fin}), which means that we diagonalize the similarity-transformed Hamiltonian, $\Bar{H}^\mathrm{ext}$, in the internal space. The external amplitudes, however, can be obtained by solving the external Schr\"odinger equation, Eq.~(\ref{eq:ext_SE}), with the conventional coupled cluster algorithm, while keeping the internal amplitudes fixed. This, of course, corresponds to the TCC model.
The new idea in this work is to iteratively solve Eqs.~(\ref{eq:int_SE_fin}) and (\ref{eq:ext_SE}) in an alternating way until self-consistency is reached. However, solving the equations exactly would not yield any computational advantage. In the next section, we discuss approximations to make this idea practical. We also describe the ecCC method, which can replace TCC as an alternative ansatz.

\subsection{Configuration interaction with iterative coupled cluster}
\label{sec:ci-icc}

In this section, we discuss approximations to solve Eqs.~(\ref{eq:int_SE_fin}) and (\ref{eq:ext_SE}). First, we always limit the excitation degree in the external space to singles and doubles in this work. That is, we rely on TCCSD to solve the external equations
\begin{equation}
    T^\mathrm{ext}_\mathrm{SD} = T^\mathrm{ext}_\mathrm{1} + T^\mathrm{ext}_\mathrm{2},
\end{equation}
which leads to the singles and doubles similarity-transformed Hamiltonian
\begin{equation}
    \Bar{H}^\mathrm{ext}_\mathrm{SD} = \e^{-T^\mathrm{ext}_\mathrm{SD}} H \e^{T^\mathrm{ext}_\mathrm{SD}}.
    \label{eq:sim_hamil_2}
\end{equation}
The exact similarity transformation of Eq.~(\ref{eq:sim_hamil_2}) yields up to six-body interactions, which would require storing up to twelve-index tensors. This is prohibitive in practice even for the smallest systems.
The cheapest reasonable truncation scheme appears to be the Baker--Campbell--Hausdorff (BCH) expansion of the similarity-transformed Hamiltonian truncated after the second order to obtain
\begin{equation}
    \Bar{H}^{\mathrm{ext},(2)}_\mathrm{SD} = \chi_0 + \sum_{pq} \chi_{pq}  \{p^\dagger q\} + \frac{1}{4} \sum_{pqrs} \chi_{pqrs}\{p^\dagger q^\dagger s r\}.
    \label{eq:2nd_quant_sim_hamil_2}
\end{equation}
The equations for obtaining the matrix elements $\chi_0$, $\chi_{pq}$, and $\chi_{pqrs}$ can be obtained with standard textbook methods\cite{Bartlett2009_Book} or, in our case, automatic equation generation tools\cite{Evangelista2022_Wicked}.
To improve upon this, we can truncate the expansion after the third order
\begin{equation}
    \Bar{H}^{\mathrm{ext},(3)}_\mathrm{SD} = \Bar{H}^{\mathrm{ext},(2)}_\mathrm{SD} + \frac{1}{(3!)^2} \sum_{pqrstu} \chi_{pqrstu}\{p^\dagger q^\dagger r^\dagger u t s\},
    \label{eq:2nd_quant_sim_hamil_3}
\end{equation}
or after the fourth order
\begin{equation}
    \Bar{H}^{\mathrm{ext},(4)}_\mathrm{SD} = \Bar{H}^{\mathrm{ext},(3)}_\mathrm{SD} + \frac{1}{(4!)^2} \sum_{pqrstuvw} \chi_{pqrstuvw}\{p^\dagger q^\dagger r^\dagger s^\dagger w v u t\}.
    \label{eq:2nd_quant_sim_hamil_4}
\end{equation}
The three- and four-body matrix elements are given in physics notation and are fully antisymmetrized. However, evaluating them in the entire space would already be prohibitive for medium-sized systems. Since the Hamiltonian is projected into the internal space, only the entries of the tensors in the active space have to be computed and stored. For comparison, the memory scaling of the truncation after the third order is that of CASPT2, where the three-body reduced density matrix is required, and the memory scaling of the truncation after the fourth order is that of NEVPT2, where four-body reduced density matrix elements are needed.

We note that, if the BCH expansion is truncated, the transformation is no longer of the type of a similarity transformation. Consequently, the eigenvalues of the transformed Hamiltonian will not exactly match those of the original Hamiltonian.

The truncation of the BCH expansion also results in dependency on the definition of the vacuum state. For example, a truncated Hamiltonian ordered with respect to the physical vacuum may have a  spectrum different from that of a normal-ordered Hamiltonian with respect to the Fermi vacuum. However, by inclusion of higher-body operators, this discrepancy will become negligible. 

In practice, we solve the following eigenvalue problem in the internal space
\begin{equation}
        (P+Q^\mathrm{int}) \Bar{H}^{\mathrm{ext},(n)}_\mathrm{SD} |\Psi_{\mathrm{int}}\rangle =  E^{(n)}_\mathrm{SD} (P+Q^\mathrm{int})  |\Psi_{\mathrm{int}}\rangle,\quad n \in \{2, 3, 4\}.
        \label{eq:int_SE_practice}
\end{equation}
Due to the structure of the CC equations, the similarity transformation is not truncated in the equations for the external amplitudes. We extract the singles and doubles amplitudes from the internal wave function in TCCSD,
\begin{equation}
    T^\mathrm{int}_\mathrm{TCCSD} = T^\mathrm{int}_\mathrm{1} + T^\mathrm{int}_\mathrm{2},
\end{equation}
and the triples and quadruples in ecCCSD,
\begin{equation}
    T^\mathrm{int}_\mathrm{ecCCSD} = T^\mathrm{int}_\mathrm{3} + T^\mathrm{int}_\mathrm{4}.
\end{equation}
In the TCCSD case, the external equations read
\begin{equation}
    Q_\mathrm{SD}^\mathrm{ext} (H  \e^{T^\mathrm{ext}_\mathrm{SD} + T^\mathrm{int}_\mathrm{TCCSD}})_C \ket{0} = 0, 
    \label{eq:tccsd_ext_SE_fin}
\end{equation}
whereas for ecCCSD, the internal and external singles and doubles are optimized in the presence of the internal triples and quadruples
\begin{equation}
    (Q_\mathrm{SD}^\mathrm{int}+Q_\mathrm{SD}^\mathrm{ext}) (H  \e^{T^\mathrm{ext}_\mathrm{SD} + T^\mathrm{int}_\mathrm{SD} + T^\mathrm{int}_\mathrm{ecCCSD}})_C \ket{0} =0.
    \label{eq:ecccsd_ext_SE_fin}
\end{equation}
The key ideas of our approach are therefore: (i) the amplitudes are optimized in the presence of dynamic correlation encoded by the external amplitudes entering Eq.~(\ref{eq:int_SE_practice}), and (ii) the external amplitudes are affected by static correlations encoded in the internal amplitudes entering Eqs.~(\ref{eq:tccsd_ext_SE_fin}) and (\ref{eq:ecccsd_ext_SE_fin}). The final wave function is an improved CCSD wave function, written as
\begin{equation}
    |\Psi_\mathrm{CASiCCSD(\textit{n})}\rangle = \e^{T^\mathrm{int}_\mathrm{SD} + T^\mathrm{ext}_\mathrm{SD}} |0\rangle.
\end{equation}
Here, we introduced the acronym CASiCCSD($n$) standing for CAS iterative CCSD. Depending on the MR-driven CC method, we refer to the approach as CASiecCCSD($n$) or CASiTCCSD($n$), where $n$ denotes the order after which the BCH expansion is truncated.
More generally, we denote the method as CASiCC.

\begin{figure}
  \centering
\includegraphics[width=0.5\textwidth]{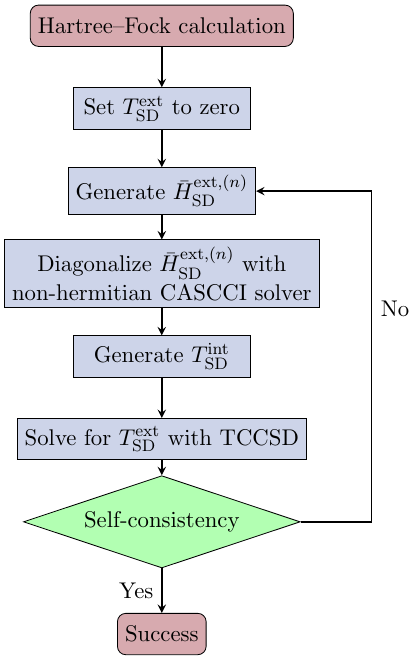}
  \caption{Flowchart for the CASiTCCSD(\textit{n}) algorithm.}
  \label{fig:icascc_procedure}
\end{figure}

To conclude this section, we summarize the CASiCC algorithm at the example of the CASiTCCSD(3) model in Fig.~\ref{fig:icascc_procedure}: first, we start with a Hartree--Fock calculation to generate the initial hermitian Hamiltonian. This corresponds to the similarity-transformed $\Bar{H}^{\mathrm{ext},(3)}_\mathrm{SD}$ with $T^\mathrm{ext}_\mathrm{SD}=0$. Then, the Hamiltonian is diagonalized within the active space which gives an initial set of CASCI 
coefficients. With the singles and doubles CI-coefficients, we generate the initial internal amplitudes $T^\mathrm{int}_\mathrm{SD}$. Subsequently, we perform a TCCSD calculation to obtain the external amplitudes $T^\mathrm{ext}_\mathrm{SD}$. Up to this point, the algorithm corresponds to the TCCSD method. Next, we generate the first nonhermitian Hamiltonian, $\Bar{H}^{\mathrm{ext},(3)}_\mathrm{SD}$, with the external amplitudes $T^\mathrm{ext}_\mathrm{SD}$. This Hamiltonian is then diagonalized to obtain a new set of CI coefficients, which are employed to generate the new internal amplitudes, $T^\mathrm{int}_\mathrm{SD}$ subjected to the TCCSD algorithm to generate a new set of external amplitudes, $T^\mathrm{ext}_\mathrm{SD}$. At this point, we check the energy difference between the current and the previous iteration. If the energy is converged, the algorithm is terminated, otherwise, $\Bar{H}^{\mathrm{ext},(3)}_\mathrm{SD}$ is calculated with the new external amplitudes $T^\mathrm{ext}_\mathrm{SD}$ and the next iteration step begins. 

\section{Computational details}
\label{sec:comp_details}

For nonhermitian $n$-body CASCI calculations, we have developed a Python program, which exploits C++ routines with pybind11\cite{pybind11} for the performance sensitive tasks. This program can diagonalize real symmetric and nonsymmetric Hamiltonians with up to four-body operators and integrates the CC routines with the DIIS algorithm\cite{Pulay1980_DIIS,Scuseria1986_DIISCC}. We relied on Wick\&d\cite{Evangelista2022_Wicked} for the automatic code generation for the CC equations, for the similarity transformation, and for the transformation of the CI coefficients to CC amplitudes.
The source code of our program, the Python scripts for reproducing the calculations, and all data presented in this work have been published on Zenodo \cite{casicc}. The convergence thresholds used for the CC calculations were an energy threshold of 10$^{-8}$~Hartree and a threshold for the norm of the amplitudes of 10$^{-5}$ atomic units.

All DMRG calculations were conducted with the QCMaquis software package\cite{QCMaquis_2015}. 
We sorted the orbitals on the lattice according to the Fiedler ordering\cite{Legeza2003_OrderingOptimization,Reiher2011_Fiedler} and converged the energy up to 10$^{-6}$~Hartree with a maximum bond dimension, denoted by $m$, of 2000.
It was shown before\cite{Chan2004_DMRG_N2} that DMRG results obtained with a bond dimension of 2000 are converged up to 0.5~mHartree for N$_2$ with the cc-pVDZ basis set. 
Since  N$_2$ is the largest system that we consider here, that bond dimension will be sufficient for all other systems analyzed in this work.
Furthermore, the accuracy of 0.5~mHartree meets our requirements for evaluating the performance of the CC methods.

The error of an electronic structure model ('$\mathrm{M}$') at some point $x$ on the potential energy curve measured against a reference model ('$\mathrm{ref}$') is evaluated as $\Delta E(x) = E_\mathrm{M}(x) - E_\mathrm{ref}(x)$.

For the atomic orbital basis, we chose Pople \cite{Pople1980_self} (for such small basis sets one can still obtain the FCI solution by direct diagonalization) and correlation consistent basis sets \cite{Dunning1989_gaussian}.
The calculation of the molecular orbital integrals in the restricted Hartree--Fock basis, CASSCF, and NEVPT2 calculations were performed with the PySCF program\cite{Sun2018_PySCF1,Sun2020_PySCF2}.

\section{Results}
\label{sec:application}

\subsection{H$_4$ and H$_8$ test systems}
\label{sec:h4_h8}

\begin{figure}
\centering
\includegraphics[width=0.18\textwidth]{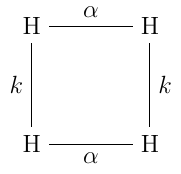}
\caption{Geometry of the stretched H$_4$ system. The $k$ parameter is fixed to 2~{\AA} and the $\alpha$ parameter is varied between 0.5 and 4~{\AA}.}
\label{fig:p4_system}
\end{figure}

To demonstrate the effectiveness of the CASiCC method, we first examine the dissociation of the  stretched H$_4$ system into two H$_2$ for a CAS(4,4). The system was first introduced by Paldus and Jankowski\cite{Paldus1980_P4} and its geometric configuration is illustrated in Fig.~\ref{fig:p4_system}, forming a perfect square. The active space was chosen based on our previous work, where we analyzed the entanglement and mutual information.\cite{Morchen2022_TCC}
In our comparative analysis of different methods, we specifically selected the stretched geometry with an edge length of $k=2\ \text{\AA}$ over the more commonly used $k=1\ \text{\AA}$. This decision was based on the observation that the former geometry yielded more pronounced differences between the methods, thereby providing a better basis for the comparison.
A key feature of this system is the adjustability of its multireference characteristics through the dissociation parameter $\alpha$. Specifically, at $\alpha = k = 2$~{\AA}, the ground state exhibits perfect degeneracy while for larger or smaller $\alpha$ values, dynamic correlation dominates. 
To analyze the capabilities of our approach, we present absolute errors in electronic energies measured against the FCI energy along the dissociation curve. This comparison comprises CCSD, TCCSD, and ecCCSD, alongside our CASiTCCSD($n$) and CASiecCCSD($n$) models.

\begin{figure}
    \centering
    \includegraphics[width=\textwidth]{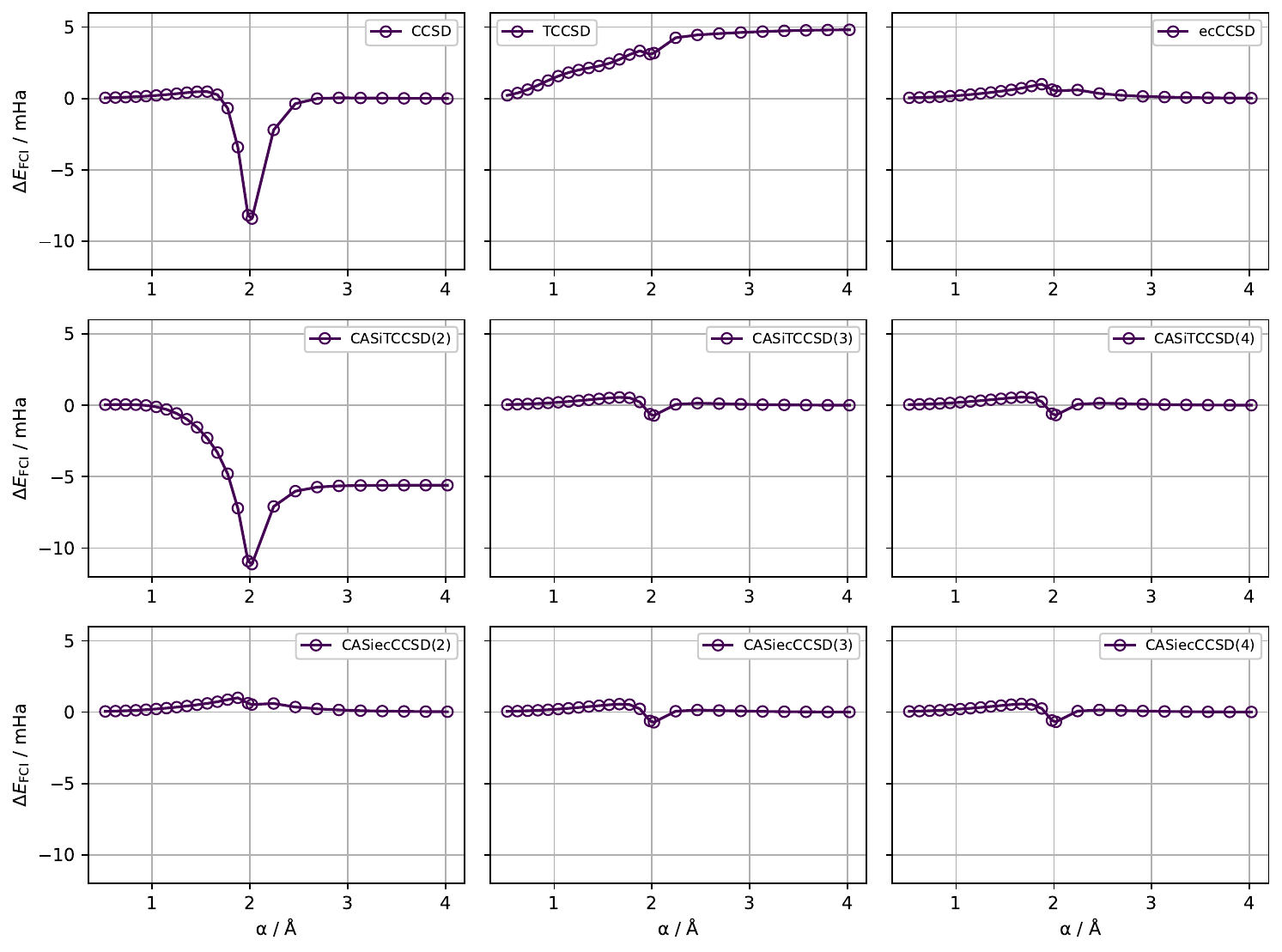}
    \caption{Error of the electronic energy in mHartree with respect to FCI for the dissociation of H$_4$ in the geometry depicted in Fig.~\ref{fig:p4_system}. Shown are results for CCSD, TCCSD, and ecCCSD (top) and for different truncation schemes of the BCH expansion for CASiTCCSD (middle) and CASiecCCSD (bottom).
    The cc-pVDZ basis set chosen yields 20 orbitals in total. The FCI calculations took all orbitals into account and we employed a CAS(4,4) for the multireference-driven CC methods.} 
    \label{fig:h4_error}
\end{figure}

Fig.~\ref{fig:h4_error} presents the values of the errors 
of various single-reference and multireference-driven CC methods with respect to FCI.
The first row of diagrams in Fig.~\ref{fig:h4_error} shows the errors of the CCSD, TCCSD, and ecCCSD energies.
CCSD fails as $\alpha$ tends to 2~{\AA} from the left and right. The error approaches 10~mHartree at 2~{\AA} and is 2 to 4 orders of magnitude larger than in the dynamic correlation regime with $\alpha<1$~{\AA} and $\alpha>3$~{\AA}.
TCCSD also shows a large systematic error reaching $>1$~mHartree for $\alpha>1$~{\AA}. In contrast to CCSD, TCCSD features no spike at 2~{\AA} and the error stays constantly high during dissociation. ecCCSD performs better than CCSD and TCCSD, but also shows a small spike at 2~{\AA} where the error is above 1~mHartree. 

The second row of Fig.~\ref{fig:h4_error} depicts the error of CASiTCCSD($n$) with truncation orders $n=2,3$ and $4$ from left to right. 
For CASiTCCSD(2), the Hamiltonian expansion has not reached convergence at the second order. This is evident as there is a significant change in the energy error curve when transitioning from $n=2$ to $n=3$. 
However, the curves for $n=3$ and $n=4$ are virtually indistinguishable, suggesting convergence with respect to $n$ has been reached for $n=3$.

The third row of Fig.~\ref{fig:h4_error} shows the different truncation schemes for CASiecCCSD($n$). The CASiecCCSD(2) results are similar to the ecCCSD results, with a slight improvement at $\alpha = 2$~{\AA}.
However, the absolute error is significantly different for $n>2$, and similar to CASiTCCSD, it converged for $n=3$ (when compared to $n=4$). Intriguingly, for both, CASiTCCSD and CASiecCCSD, the results are identical if $n>2$.
This indicates that our approach of accounting for static correlation, independently of whether through the inclusion of singles and doubles or through the inclusion of triples and quadruples from the CAS, does not affect the energy when applied in a self-consistent manner.

From a computational point of view, the TCCSD is significantly more appealing. Specifically, ecCCSD necessitates the calculation of $T_3$ and $T_4$ amplitudes, which entail memory scalings of $\mathcal{O}(O^3V^3)$ and $\mathcal{O}(O^4V^4)$, respectively, where $O$ and $V$ represent the numbers of occupied and virtual orbitals within the active space, respectively. 
Moreover, the key challenges in the implementation of the (parent) ecCCSD
are: (i) the transformation of CI coefficients to CC amplitudes is an issue when efficiency is a priority, and (ii) efficiently integrating the contributions of triples and quadruples into the singles and doubles residuals is equally challenging.

Given these considerations, the convergence of both, CASiTCCSD and CASiecCCSD, to identical results favors the CASiTCCSD model in routine applications for feasibility reasons. CASiTCCSD circumvents the aforementioned computational complexities, making it a more efficient and practical choice.

\begin{table}[ht]
\centering
\begin{tabular}{lccc}
\hline\hline
           &CCSD   &          TCCSD   &         ecCCSD   \\ 
\hline
H$_4$      &8.41   &           4.59   &           0.97    \\ 
H$_8$      &9.05   &           2.17   &           5.77   \\ 
\hline
           &    CASiTCCSD(2)   &    CASiTCCSD(3) &    CASiTCCSD(4)   \\ 
\hline 
H$_4$      &          11.12   &           0.71   &           0.68   \\ 
H$_8$      &           6.00   &           6.82   &           6.83   \\ 
\hline
           &   CASiecCCSD(2)   &   CASiecCCSD(3)   &   CASiecCCSD(4)  \\
\hline
H$_4$      &           0.97   &           0.71   &           0.68  \\
H$_8$      &           6.76   &           6.82   &           6.83  \\
\hline\hline
\end{tabular}
\caption{Nonparallelity error (NPE) 
in mHartree with respect to FCI for the dissociation of the H$_4$ and H$_8$ systems with the geometries depicted in Figs.~\ref{fig:p4_system} and \ref{fig:h8_system}, respectively. For H$_4$, we chose the cc-pVDZ basis set and a CAS(4,4) for the multireference-driven CC methods, whereas for H$_8$, we chose the 6-31G basis set and a CAS(8,8). As a reference for H$_8$, the NEVPT2-NPE evaluates to 2.14~mHa.}
\label{table:npe}
\end{table}

Next, we focused on assessing the accuracy of the CASiCC models. Fig.~\ref{fig:h4_error} shows that, with the exception of CASiTCCSD(2), the CASiCC methods pose a systematic improvement over the CCSD method in the single and also multireference regime and they overall show the lowest error, also compared to TCCSD and ecCCSD.
The nonparallelity errors (NPEs), that is the difference between the maximum and minimum error of the potential energy curves, are given in Table~\ref{table:npe}. 
According to the table, the accuracy of the models increases according to CCSD, TCCSD, and ecCCSD followed by CASiTCCSD (except for $n=2$) and CASiecCCSD which have the same errors. 
Our findings also highlight a limitation of the standard TCCSD method, revealing that it does not consistently offer an improvement over the CCSD method.

\begin{figure}
\centering
\includegraphics[width=0.29\textwidth]{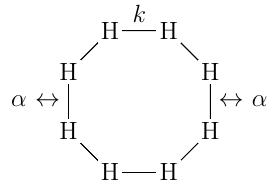}
\caption{Geometry of the H$_8$ system. The $k$ parameter denotes the bond distances for the H$_2$ units (left, right, upper, and lower ones), which are fixed to 1~{\AA}. The $\alpha$ parameters refers to shifting the left and right H$_2$ units and is varied between $-0.22$ and 0.22~{\AA}, where $\alpha=0$ corresponds to the octagon.}
\label{fig:h8_system}
\end{figure}

\begin{figure}
    \centering
    \includegraphics[width=0.6\textwidth]{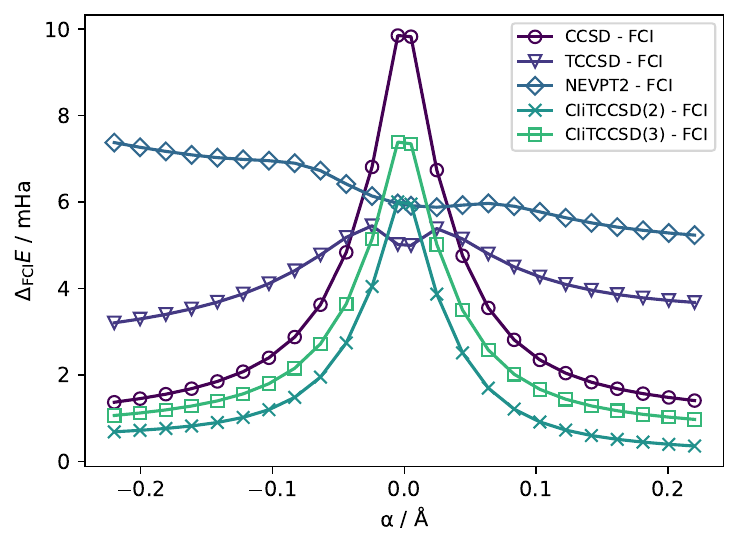}
    \caption{Error 
    of the total electronic energy in Hartree with respect to FCI 
    for the dissociation of the H$_8$ system in the geometry depicted in Fig.~\ref{fig:h8_system}. Shown are results for CCSD, TCCSD, and NEVPT2 (with CASSCF orbitals), and for CASiTCCSD for different truncation schemes of the BCH expansion.
    We chose the small 6-31G basis set for the calculation which yields 16 orbitals, for which FCI results, taking all orbitals into account, 
    can be obtained as reference. For the CASiCC calculations we employed a CAS(8,8).
    }
    \label{fig:h8_error}
\end{figure}

Continuing our analysis, we turn to the octagonal H$_8$ model system, first introduced by Piecuch and Adamowicz\cite{Piecuch1994_SplitAmplitude2_H8}, as illustrated in Fig.~\ref{fig:h8_system} with a CAS(8,8). The active space was again selected based on our previous studies based on quantum information theory measures.\cite{Morchen2022_TCC}
As for the H$_4$ system, the multiconfigurational character of H$_8$ is adjustable through the parameter $\alpha$. Our results reveal that, in accordance with our findings for H$_4$, the CASiCC models exhibit convergence at truncation order $n=3$ for H$_8$. Moreover both CASiTCCSD and CASiecCCSD again yield identical results upon convergence, as shown in Tab.~\ref{table:npe}. 

Fig.~\ref{fig:h8_error} presents the errors in total electronic energy (compared to FCI results) 
for various models: CCSD, TCCSD, and NEVPT2 with CASSCF orbitals, CASiTCCSD(2), and CASiTCCSD(3). Similar to the trends observed for H$_4$, CCSD performs well in the single-reference regime, but exhibits a pronounced peak in error at the point of degeneracy, $\alpha=0$. TCCSD also shows again a rather constant relatively large error along the entire curve, but displays only a minor peak at $\alpha=0$. CASiTCCSD again demonstrates systematic improvement over CCSD, yet the error at $\alpha=0$ is larger than that of TCCSD. The error of CASiTCCSD(2) consistently remains lower than that of CASiTCCSD(3), possibly due to fortuitous error cancellation. 
Since NEVPT2 remains one of the most widely used multireference methods, we included the results as well. The NEVPT2 curve shows the smallest peak around $\alpha=0$, but the error decreases constantly.

The NPEs of these methods are listed in Table~\ref{table:npe}. Here, CCSD shows the highest NPE, followed by the CASiCC methods and ecCCSD, while TCCSD exhibits the lowest NPE of the CC methods, and NEVPT2 yields the lowest NPE overall. Also from these results, it is evident that the CASiCC methods provide a systematic improvement over CCSD. However, TCCSD and ecCCSD might offer better performance, potentially due to error cancellation.

\subsection{Potential energy curves of small molecules}
\label{sec:h2o_n2}

For the small molecules studied in this section, we note that the iterative algorithm is usually converged to $10^{-10}$ Hartree in 5 to 15 macroiteration steps. Regarding the CC optimization of the external amplitudes, we employed the amplitudes from the previous iteration as a starting guess, which significantly accelerated convergence. The CC equations are usually converged in 2 to 3 iterations after the first few macroiterations.

\subsubsection{Dissociation of N$_2$}

\begin{figure}
    \centering
    \includegraphics[width=0.6\textwidth]{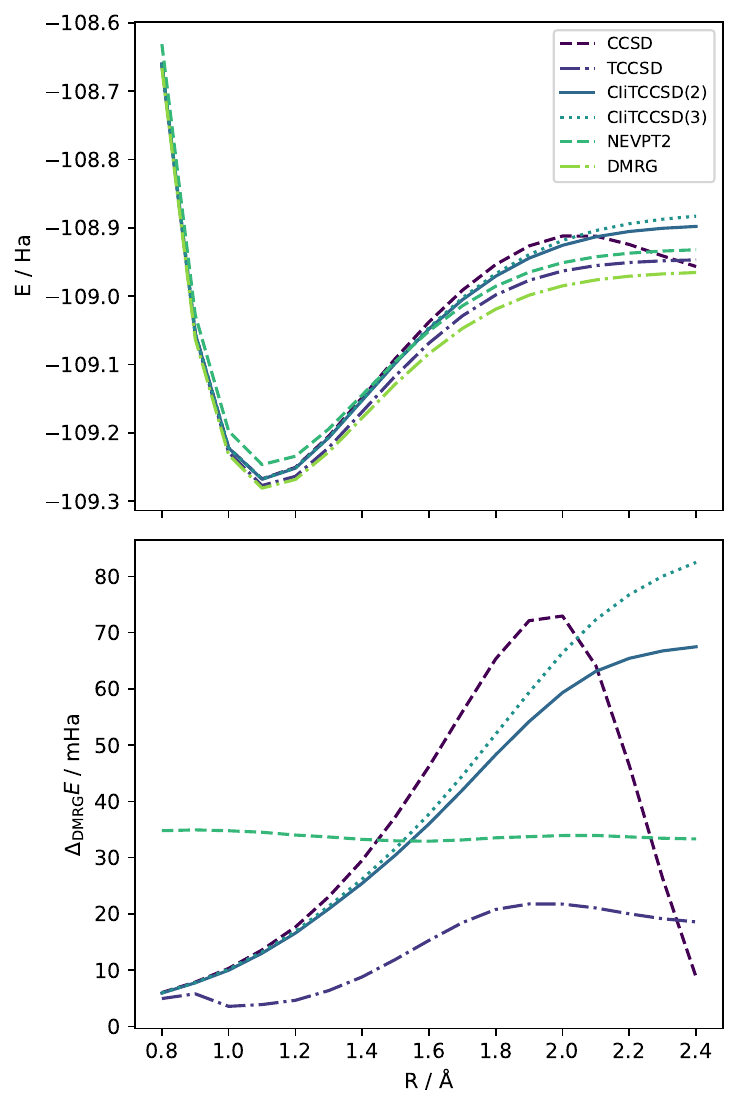}
    \caption{Total electronic energy (top) and energy error 
    with respect to DMRG($m=2000$) (bottom) for the dissociation of N$_2$ in Hartree. Shown are results for CCSD, TCCSD, and NEVPT2 (with CASSCF orbitals), and for CASiTCCSD for different truncation schemes of the BCH expansion.
    The cc-pVDZ basis set was chosen, which yielded 28 orbitals. For the DMRG reference calculation, all 28 orbitals were all taken into account in order to obtain an energy close to the exact FCI reference. All other results were obtained in a limited orbital space of CAS(6,6).}
    \label{fig:n2_error}
\end{figure}

To describe the bond-breaking in N$_2$, we chose a CAS(6,6). The cc-pVDZ basis set yielded a total of 28 orbitals.
Potential energy curves for CCSD, NEVPT2 with CASSCF orbitals, CASiTCCSD(2), CASiTCCSD(3), TCCSD, and DMRG (top figure), along with the deviation of the MRPT and coupled cluster (CC) models compared to the DMRG (bottom figure), are shown in Fig.~\ref{fig:n2_error}.

As expected, CCSD completely fails in the dissociation limit, whereas TCCSD reliably describes the dissociation of N$_2$, in agreement with initial results for TCC\cite{Kinoshita2005_TailoredCC1}. The CASiTCCSD curves for $n=2$ and $n=3$ are qualitatively correct, but exhibit a substantial error at the dissociation limit. 
We ascribe these shortcomings to two main sources of error: (i) CASiCC is not a genuine multireference method; as a multireference-driven single-reference method, 
there is a persistent asymmetry in the excitations due to the preference of the reference determinant
in the case of near or exact degeneracies. (ii) Additionally, correlation in the external space increases as well upon dissociation so that higher cluster amplitudes must be included to account for this correlation. To be more specific, Piecuch, Kucharski, and Bartlett observed that at least triple and quadruple contributions with semi-internal excitations have to be incorporated to account for this correlation.\cite{Piecuch1999_CCSDt}
Hence, the integration of higher-order cluster amplitudes into the external space becomes a critical consideration for the future development of the CASiTCCSD method.
Note again that CASiTCCSD(3) and CASiecCCSD(3) yielded virtually the same result, however, ecCCSD converged significantly slower in the strong correlation limit.
Lastly, we highlight that the NEVPT2 method is significantly more accurate than all CC models, although it exhibits a systematic deviation. 

\begin{figure}
    \centering
    \includegraphics[width=0.6\textwidth]{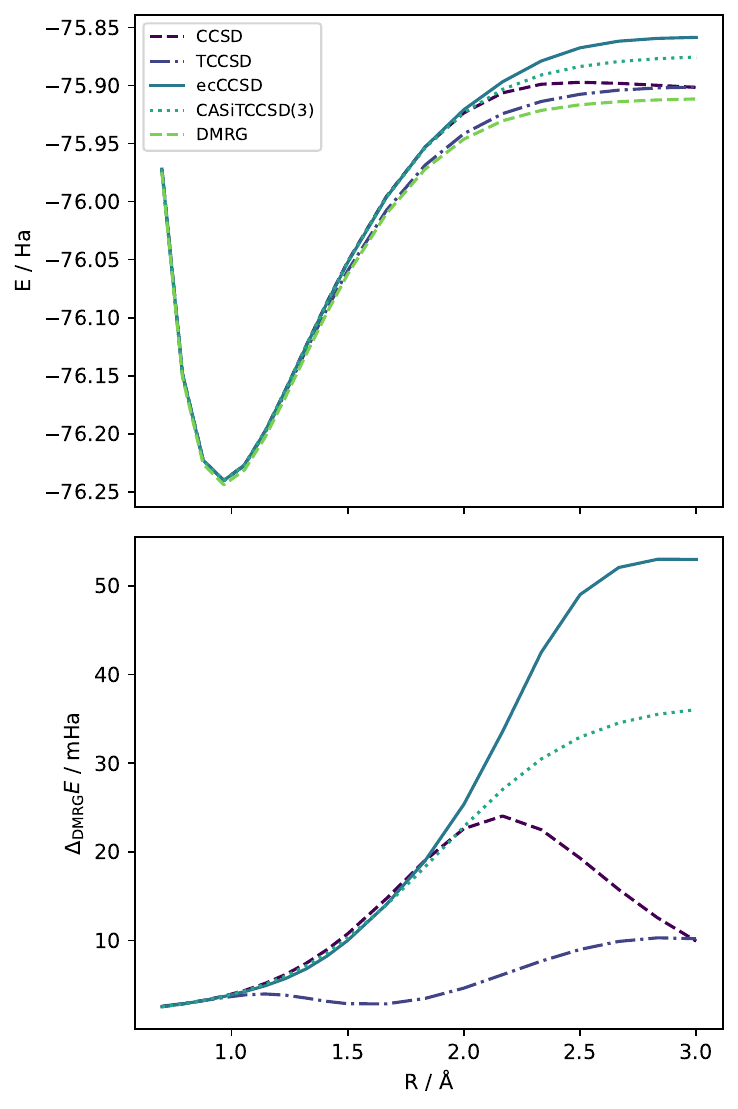}
    \caption{Total electronic energy (top) and energy error (bottom) with respect to DMRG($m=2000$) for the dissociation of H$_2$O in Hartree. Shown are the results of CCSD, TCCSD, ecCCSD, and CASiTCCSD(3) calculations.
We chose the cc-pVDZ basis set, which gives 24 orbitals in total. For the DMRG calculations, these were all taken into account. Hence, the DMRG results are expected to be very close to the FCI result. By contrast, all other results were based on a CAS(6,5) reference.}
    \label{fig:h2o_error}
\end{figure}

\subsubsection{Dissociation of H$_2$O}
Finally, we examine the symmetric double dissociation of the H$_2$O molecule with a bond angle of $104.5^\circ$ and a CAS(6,5). The dissociation curves for CCSD, TCCSD, ecCCSD, CASiTCCSD(3), and DMRG, along with the energy errors of the CC methods relative to those of DMRG, are provided in Fig.~\ref{fig:h2o_error}.
H$_2$O also becomes strongly correlated in the dissociation limit, which leads again to the  unphysical results of CCSD. TCCSD, by contrast, correctly dissociates H$_2$O with an error in the dissociation limit of roughly 10~mHartree. 
ecCCSD describes the dissociation qualitatively well, but it is affected by a large error of around 50~mHartree which has already been discussed in the literature\cite{Chan2021_ecCCSD-DMRG}. CASiTCCSD(3) also cures the deficiencies of CCSD and correctly dissociates the molecule, with an error of around 35~mHartree, 
which is between TCCSD and ecCCSD. 

Hence, again we find that CASiTCCSD(3) delivers a systematic improvement over the CCSD results. Nonetheless, for H$_2$O and N$_2$ TCCSD  shows higher accuracy in the limit of strong correlation. It was also observed recently\cite{Pittner2023_HilbertSpaceTCC} that, due to error compensation, single-reference TCC yields results similar to the more rigorous Hilbert-space multireference TCC approach even for very strongly correlated systems, also pointing to an error compensation mechanism. 
The error of CASiCC in the strong correlation regime can be ascribed to the missing correlation in the external space, and to the bias of the reference determinant.

\section{Conclusions}
\label{sec:conclusions}

In this study, we introduced a configuration interaction approach with iterative coupled cluster feedback (CASiCC). CASiCC iteratively combines CASCI self-consistently with multireference-driven coupled cluster methods. 
The iterative feedback loop is achieved  by similarity transforming the Hamiltonian of the CASCI calculation with the external amplitudes.
The similarity transformation results in a nonhermitian Hamiltonian with up to $n$-body terms. The nonhermiticity does not pose any technical problem, because available CI programs can be easily adapted, and we have demonstrated that the expansion of the Hamiltonian converges already at the third order.
Moreover, we showed that the results of CASiCC are quantitatively indistinguishable for the optimization of the amplitudes with tailored coupled cluster from externally corrected coupled cluster. 
This fact strongly favors the CASiTCC variant, since its implementation and computational scaling are more advantageous, with
CASiTCCSD(3) as the most promising candidate.

We studied CASiCC for dissociation curves of H$_4$, H$_8$, N$_2$, and H$_2$O. Notably, during the dissociation of N$_2$ and H$_2$O, we observed that TCCSD exhibits superior performance, which may be attributed to the fortuitous error cancellation. 
Our reason for assuming this fortuitous error cancellation is based on the fact that in the weakly-correlated regime, the TCCSD method may be significantly worse compared to CCSD, but in the strongly-correlated regime TCCSD removes a significant amount of the error made by CCSD. However, a method that is not based on error compensation should be able to systematically improve upon CCSD in both regimes. 
For further refinement of the CASiCC method, the effect of higher excitations in the external space should be investigated. 

However, given the fact that we could not systematically improve the TCCSD method iteratively suggests that the accuracy of multireference driven CC methods is inherently limited by the reference bias. 

To maintain cost efficiency similar to that of the standard CCSD method, we are exploring the integration of the DMRG algorithm as a CASCI solver. This approach will build on our recent work \cite{Baiardi2020_Transcorrelated1, Baiardi2022_Transcorrelated2}, where we have employed DMRG to evaluate ground state energies of 
nonhermitian Hamiltonians with up to three-body operators.

\section*{Acknowledgments}

M.~R. gratefully acknowledges financial support by the Swiss National Science Foundation (grant no. 200021\_219616), by ETH Zurich (through ETH grant ETH-43 20-2), and by the `Quantum for Life Center'' funded by the Novo Nordisk Foundation (grant NNF20OC0059939). R.~F. is grateful for a PhD fellowship awarded by the G\"unthard Foundation. J.~L. and M.~L. acknowledge the support from the European Partnership on Metrology (project number 22IEM04 MQB-Pascal), cofinanced from the European Union’s Horizon Europe Research and Innovation Programme and by the Participating States.

%\bibliography{bibliography}
\providecommand{\latin}[1]{#1}
\makeatletter
\providecommand{\doi}
  {\begingroup\let\do\@makeother\dospecials
  \catcode`\{=1 \catcode`\}=2 \doi@aux}
\providecommand{\doi@aux}[1]{\endgroup\texttt{#1}}
\makeatother
\providecommand*\mcitethebibliography{\thebibliography}
\csname @ifundefined\endcsname{endmcitethebibliography}
  {\let\endmcitethebibliography\endthebibliography}{}

\end{document}